\documentclass[12pt]{article}

\begin{document}

\title{P, C and T: Different Properties on the Kinematical Level}
\author{Valeriy V. Dvoeglazov\\
UAF, Universidad Aut\'onoma de Zacatecas \\
Ap. Postal 636, Suc. 3, Zacatecas 98061, Zac., M\'exico\\
E-mail: valeri@fisica.uaz.edu.mx
}

\date{\empty}


 
\maketitle

\begin{abstract}
\small{We study the discrete symmetries ($P$,$C$ and $T$) on 
the kinematical level within the extended Poincar\'e Group.
On the basis of the Silagadze research, we  investigate the question of the definitions of the discrete symmetry operators 
both on the classical level, and in the secondary-quantization  scheme. We study the physical contents 
within several bases: light-front formulation,  helicity basis, angular  momentum basis, and so on, on several practical examples.
We analize problems in construction of the neutral particles in the the  $(1/2,0)+(0,1/2)$ representation, the 
$(1,0)+(0,1)$ and the $(1/2,1/2)$ representations of the Lorentz Group. As well known, the photon has the quantum numbers $1^-$, so the $(1,0)+(0,1)$ representation of the Lorentz group is relevant to its description. We have ambiguities in the definitions of the corresponding operators $P$, $C$; $T$, which lead to different physical consequences. It appears that the answers are connected with the helicity basis properties, and 
commutations/anticommutations of the corresponding operators, $P$, $C$, $T$, and $C^2$, $P^2$, $(CP)^2$ properties.} 
\end{abstract}



\section{Introduction.}

In his paper of 1992 Silagadze claimed: ``It is shown that the usual situation when boson and its antiparticle have the same internal parity, while, fermion and its antiparticle have opposite parities, assumes a kind of locality of the theory. In general, when a quantum-mechanical parity operator is defined by means of the group extension technique, the reversed situation is also possible", Ref.~\cite{Silagadze}.
Then, Ahluwalia {\it et al} proposed~\cite{AHLU} the
Bargmann-Wightman-Wigner- type quantum field theory,
where, as they claimed, the boson and the antiboson have oposite intrinsic parities (see also~\cite{Dva}).
Actually, this type of theories has been first proposed by Gelfand and
Tsetlin (1956), Ref.~\cite{Gelfand}. In fact, it is based on the two-dimensional representation
of the inversion group. They
indicated applicability of this theory to the description of the system of
$K$-mesons and the possible relations to the Lee-Yang paper.
The (anti)comutativity of the discrete symmetry operations has
also been investigated by Foldy and Nigam~\cite{Nigam}, who claimed that 
it is related to the eigenvalues of the charge operator. The relations of
the Gelfand-Tsetlin construct to the representations of the
anti-de Sitter $SO(3,2)$ group and the general relativity theory
have also been discussed
in subsequent papers of Sokolik. E. Wigner~\cite{Wig}
presented some relevant results at the Istanbul School
on Theoretical Physics in 1962. Later, Fushchich {\it et al} discussed
the wave equations.  Actually, the theory presented
by Ahluwalia, Goldman and Johnson is
the Dirac-like generalization of the Weinberg $2(2J+1)$-theory
for the spin 1. The equations have also been presented in the Sankaranarayanan
and Good paper of 1965, Ref.~\cite{SG}. 
In~\cite{DasG} the theory in the $({1\over
2},0)\oplus (0,{1\over 2})$ representation based on
the chiral helicity 4-eigenspinors was proposed.  The corresponding equations
have been obtained, e.~g., in~\cite{DVO2}.
However, later we found the papers by Ziino and Barut~\cite{Barut}
and the Markov papers~\cite{Markov}, which also have connections
with the subject under consideration. 
The question of definitions of the discrete symmetries operators raised by Silagadze,
has not yet been clarified in detail. Explicit examples are presented below and in the previous 
papers~\cite{DVO1,DVO2, DVALFP,AHLU,Dva,SG,DasG,Barut,Markov}. 


\section{Helicity Basis and Parity.}       

The 4-spinors have been studied  well
when the basis has been chosen in such a way that they are eigenstates
of the $\hat {\bf S}_3$ operator:
\begin{eqnarray}
&&u_{{1\over 2},{1\over 2}} = N_{1\over 2}^+ \begin{pmatrix}{1\cr0\cr1\cr0\cr}\end{pmatrix},
u_{{1\over 2},-{1\over 2}} =N_{-{1\over 2}}^+ \begin{pmatrix}{0\cr1\cr0\cr1\cr}
\end{pmatrix},\label{sb1a}\\
&&v_{{1\over 2},{1\over 2}} = N_{1\over 2}^-\begin{pmatrix}{1\cr0\cr-1\cr0\cr}\end{pmatrix},
v_{{1\over 2},-{1\over 2}} =N_{-{1\over 2}}^-
\begin{pmatrix}{0\cr1\cr0\cr-1\cr}\end{pmatrix}.\label{sb1}
\end{eqnarray}

And, oppositely, the
helicity basis case has not been studied almost at all
(see, however, Refs.~\cite{Novozh,JW}).  Let me remind
that the boosted 4-spinors $\Psi ({\bf p}) = column\,  (\phi_R ({\bf p})\,\, \pm \phi_L ({\bf p}))$ in 
the common-used basis are 
the parity eigenstates with the eigenvalues of $\pm 1$. 

In the helicity spin basis the 2-eigenspinors of the helicity operator~\cite{Var}
\begin{eqnarray}
{1\over 2} {\bf  \sigma}\cdot\widehat
{\bf p} = {1\over 2} \begin{pmatrix}{\cos\theta & \sin\theta e^{-i\phi}\cr
\sin\theta e^{+i\phi} & - \cos\theta\cr}\end{pmatrix},
\end{eqnarray}
$\theta , \phi$ are the angles of the spherical coordinate system,
can be defined as follows~\cite{Var,Dv1}:
\begin{eqnarray}
\phi_{{1\over 2}\uparrow}\sim\begin{pmatrix}{\cos{\theta \over 2} e^{-i\phi/2}\cr
\sin{\theta \over 2} e^{+i\phi/2}\cr}\end{pmatrix}\,,\quad
\phi_{{1\over 2}\downarrow}\sim\begin{pmatrix}{\sin{\theta \over 2} e^{-i\phi/2}\cr
-\cos{\theta \over 2} e^{+i\phi/2}\cr}\end{pmatrix}\,,\quad\label{ds}
\end{eqnarray}
for $h=\pm 1/2=\uparrow\downarrow$ eigenvalues, respectively.

We start from the Klein-Gordon equation, generalized for
describing the spin-1/2  particles (i.~e., two degrees
of freedom), $c=\hbar=1$:
\begin{equation}
(E+{\bf  \sigma}\cdot {\bf p}) (E- {\bf  \sigma}\cdot {\bf p}) \phi
= m^2 \phi\,.\label{de}
\end{equation}
It can be re-written in the form of the system of two first-order equations
for 2-spinors as in the Sakurai book. At the same time, we observe that they may be chosen as the eigenstates of the helicity operator:
\begin{eqnarray}
(E-({\bf \sigma}\cdot {\bf p})) \phi_\uparrow &=& (E-p) \phi_\uparrow
=m\chi_\uparrow \,,\\
(E+({\bf \sigma}\cdot {\bf p})) \chi_\uparrow &=& (E+p) \chi_\uparrow
=m\phi_\uparrow \,,
\end{eqnarray}

\begin{eqnarray}
(E-({\bf \sigma}\cdot {\bf p})) \phi_\downarrow &=& (E+p) \phi_\downarrow
=m\chi_\downarrow\,, \\
(E+({\bf \sigma}\cdot {\bf p})) \chi_\downarrow &=& (E-p) \chi_\downarrow
=m\phi_\downarrow \,.
\end{eqnarray}
If the $\phi$ spinors are defined by the equation (\ref{ds}) then we
can construct the corresponding $u-$ and $v-$ 4-spinors\footnote{One can
also try to construct yet another theory differing from
the ordinary Dirac theory. The 4-spinors may be {\it not}
the eigenspinors of the helicity operator of the $(1/2,0)\oplus (0,1/2)$
representation space, cf.~\cite{DasG}. They might be
the eigenstates of the {\it chiral} helicity operator introduced
in~\cite{DasG}.} 
\begin{eqnarray}
u_\uparrow &=&
N_\uparrow^+ \begin{pmatrix}{\phi_\uparrow\cr {E-p\over m}\phi_\uparrow\cr}\end{pmatrix} =
{1\over \sqrt{2}}\begin{pmatrix}{\sqrt{{E+p\over m}} \phi_\uparrow\cr
\sqrt{{m\over E+p}} \phi_\uparrow\cr}\end{pmatrix},\label{s1a}\\
u_\downarrow &=& N_\downarrow^+ \begin{pmatrix}{\phi_\downarrow\cr
{E+p\over m}\phi_\downarrow\cr}\end{pmatrix} = {1\over
\sqrt{2}}\begin{pmatrix}{\sqrt{{m\over E+p}} \phi_\downarrow\cr \sqrt{{E+p\over
m}} \phi_\downarrow\cr}\end{pmatrix},\label{s1}\\
v_\uparrow &=& N_\uparrow^- \begin{pmatrix}{\phi_\uparrow\cr
-{E-p\over m}\phi_\uparrow\cr}\end{pmatrix} = {1\over \sqrt{2}}\begin{pmatrix}{\sqrt{{E+p\over
m}} \phi_\uparrow\cr
-\sqrt{{m\over E+p}} \phi_\uparrow\cr}\end{pmatrix},\label{s2a}\\
v_\downarrow &=& N_\downarrow^- \begin{pmatrix}{\phi_\downarrow\cr
-{E+p\over m}\phi_\downarrow\cr}\end{pmatrix} = {1\over
\sqrt{2}}\begin{pmatrix}{\sqrt{{m\over E+p}} \phi_\downarrow\cr -\sqrt{{E+p\over
m}} \phi_\downarrow\cr}\end{pmatrix},\label{s2}
\end{eqnarray}  
where the normalization to the unit ($\pm 1$) was used.
One can prove that the matrix
$P=\gamma^0 = \begin{pmatrix}{0&I\cr I & 0\cr}\end{pmatrix}$
can be used in the parity operator as 
in the original Dirac basis. Indeed, the 4-spinors
(\ref{s1a},\ref{s2}) satisfy the Dirac equation in the spinorial (Weyl)
representation of the $\gamma$-matrices. Hence, the parity-transformed function
$\Psi^\prime (t, -{\bf x})=P\Psi (t,{\bf x})$ must satisfy
$[i\gamma^\mu \partial_\mu^{\,\prime} -m ] \Psi^\prime (t,-{\bf x}) =0$
with $\partial_\mu^{\,\prime} = (\partial/\partial t, -{\bf \nabla}_i)$.
This is possible when $P^{-1}\gamma^0 P = \gamma^0$ and
$P^{-1} \gamma^i P = -\gamma^i$. The P-matrix above
satisfies these requirements, as in the textbook case~\cite{Itzyk}.

Next, it is easy to prove that one can form the projection
operators 
\begin{eqnarray}
{\cal P}_+ &=& +\sum_{h} u_h ({\bf p}) \bar u_h ({\bf p})
=\frac{p_\mu \gamma^\mu +m}{2m},\,\,\\
{\cal P}_- &=& -\sum_{h} v_h ({\bf p}) \bar v_h ({\bf p})
= \frac{m- p_\mu \gamma^\mu}{2m},
\end{eqnarray} 
with the properties $P_+ +P_- =1$ and $P_\pm^2 = P_\pm$.
This permits us to expand the 4-spinors defined in the basis  (\ref{sb1a},\ref{sb1})
in  the linear superpositions of the helicity basis
4-spinors, and to find corresponding coefficients of the expansion: 
\begin{equation}
u_\sigma ({\bf p}) = A_{\sigma h} u_h ({\bf p})
+ B_{\sigma h} v_h ({\bf p}),\,\,
v_\sigma ({\bf p}) = C_{\sigma h} u_h ({\bf p})
+ D_{\sigma h} v_h ({\bf p}).
\end{equation} 
Multiplying the above equations by $\bar u_{h^\prime}$,
$\bar v_{h^\prime}$ respectively, and using the normalization conditions,
we obtain $A_{\sigma h }= D_{\sigma h}= \bar u_h u_\sigma$,
$B_{\sigma h}=C_{\sigma h}= - \bar v_h u_\sigma$.
Thus,  the transformation matrix from the common-used basis to the helicity basis is
\begin{equation}
\begin{pmatrix}{u_\sigma\cr
v_\sigma\cr}\end{pmatrix}={\cal U} \begin{pmatrix}{u_h\cr
v_h\cr}\end{pmatrix},\,\,\quad{\cal U} = \begin{pmatrix}{A&B\cr
B&A}\end{pmatrix}
\end{equation}
Neither $A$ nor $B$ are unitary: 
\begin{eqnarray}
A= (a_{++} +a_{+-}) (\sigma_\mu a^\mu) +(-a_{-+} +a_{--})
(\sigma_\mu a^\mu) \sigma_3\,,\\
B= (-a_{++} +a_{+-}) (\sigma_\mu a^\mu) +(a_{-+} +a_{--})
(\sigma_\mu a^\mu) \sigma_3\,,
\end{eqnarray}
where 
\begin{eqnarray}
a^0 &=& -i\cos (\theta/2) \sin (\phi/2) \in \Im m\,, \,\, a^1 = \sin (\theta/2) \cos (\phi/2)\in \Re e,\nonumber\\
a^2 &=& \sin (\theta/2) \sin (\phi/2) \in \Re e\,,\,\,
a^3 = \cos (\theta/2) \cos (\phi/2)\in \Re e,
\end{eqnarray}
and
\begin{eqnarray}
a_{++} &=&\frac{\sqrt{(E+m)(E+p)}}{2\sqrt{2} m}\,,\,
a_{+-} =\frac{\sqrt{(E+m)(E-p)}}{2\sqrt{2} m}\,,\\
a_{-+} &=&\frac{\sqrt{(E-m)(E+p)}}{2\sqrt{2} m}\,,\,
a_{--} =\frac{\sqrt{(E-m)(E-p)}}{2\sqrt{2} m}\,.
\end{eqnarray}
However, $A^\dagger A+B^\dagger B =I$, so the matrix ${\cal U}$
is unitary. Please note that this matrix acts
on the {\it spin}  indices ($\sigma$, $h$), and not on
the spinorial indices; it is $4\times 4$ matrix.  

We now investigate the properties of the helicity-basis 4-spinors
with respect to the discrete symmetry operations $P,C$ and $T$.
It is expected that $h\rightarrow -h$ under parity,
as Berestetski\u{\i}, Lifshitz and Pitaevski\u{\i}
claimed~\cite{Lan}. Indeed, if ${\bf x}\rightarrow -{\bf x}$,
then the vector ${\bf p}\rightarrow -{\bf p}$, but the axial vector
${\bf S}\rightarrow {\bf S}$, that implies the above statement.
The helicity 2-eigenspinors
transform $\phi_{\uparrow\downarrow} \Rightarrow
-i \phi_{\downarrow\uparrow}$ with respect to ${\bf p} \rightarrow -{\bf p}$, Ref.~\cite{Dv1}.
Hence, 
\begin{eqnarray}
Pu_\uparrow (-{\bf p}) &=& -i u_\downarrow ({\bf p})\,,
Pv_\uparrow (-{\bf p}) = +i v_\downarrow ({\bf p})\,,\\
Pu_\downarrow (-{\bf p}) &=& -i u_\uparrow ({\bf p})\,,
Pv_\downarrow (-{\bf p}) = +i v_\uparrow ({\bf p})\,.
\end{eqnarray} 
Thus, on the level of classical fields, we observe that
the helicity 4-spinors transform to the 4-spinors of the opposite
helicity.

Also,
\begin{eqnarray}
Cu_\uparrow ({\bf p}) &=& - v_\downarrow ({\bf p})\,,
Cv_\uparrow ({\bf p}) = +  u_\downarrow ({\bf p})\,,\\
Cu_\downarrow ({\bf p}) &=& + v_\uparrow ({\bf p})\,,
Cv_\downarrow ({\bf p}) = - u_\uparrow ({\bf p})\,.
\end{eqnarray}
due to the properties of the Wigner operator $\Theta \phi_\uparrow^\ast =
-\phi_\downarrow$ and $\Theta \phi_\downarrow^\ast = +\phi_\uparrow$, $\Theta_{[1/2]} = -i\sigma_2$.
Similar conclusions can be obtained  in the Fock space.

We define the field operator as follows:
\begin{equation}
\Psi (x^\mu) = \sum_h \int \frac{d^3 {\bf p}}{(2\pi)^3}
\frac{\sqrt{m}}{2E} [ u_h ({\bf p}) a_h ({\bf p}) e^{-ip_\mu x^\mu} +v_h ({\bf p})
b^\dagger_h ({\bf p}) e^{+ip_\mu x^\mu} ]\,.
\end{equation}
The commutation
relations are assumed to be the standard
ones~\cite{Bogol,Wein,Itzyk,Greib} (compare  with Refs.~\cite{DVO2,DasG}). 
If one defines $U_P \Psi (x^\mu) U_P^{-1} = \gamma^0  \Psi
(x^{\mu^\prime})$, $U_C \Psi (x^\mu) U_C^{-1} =  C \Psi^\dagger (x^\mu)$
and the anti-unitary operator of time reversal $(V_T \Psi (x^\mu)
V_T^{-1})^\dagger = T \Psi^\dagger (x^{\mu^{\prime\prime}})$,
then it is easy to obtain the corresponding transformations
of the creation/annihilation operators: 
\begin{eqnarray}
&&U_P a_h ({\bf p}) U_P^{-1} = -i a_{-h} (-{\bf p}),\,\,\,
U_P b_h ({\bf p}) U_P^{-1} = -i b_{-h} (-{\bf p}),\label{pa1}\\
&&U_C a_h ({\bf p}) U_C^{-1} = (-1)^{{1\over 2}+h} b_{-h} ({\bf
p}), \,\,\,U_C b_h ({\bf p}) U_C^{-1} = (-1)^{{1\over 2}-h} a_{-h}
(-{\bf p}).\nonumber\\
\end{eqnarray} 
As a consequence, we obtain (provided that $U_P \vert 0>=\vert 0>$,
$U_C\vert 0>= \vert 0>$)
 \begin{eqnarray}
&&U_P a^\dagger_h ({\bf p}) \vert 0>=
U_P a_h^\dagger U_P^{-1} \vert 0> =i a_{-h}^\dagger
(-{\bf p}) \vert 0>= i \vert -{\bf p}, -h >^+\,,\nonumber\\ 
&&\\
&&U_P
b^\dagger_h ({\bf p}) \vert 0> =U_P b_h^\dagger U_P^{-1}
\vert 0> =i b_{-h}^\dagger (-{\bf p}) \vert 0>= i \vert -{\bf p},
-h >^-\,,\nonumber\\
\end{eqnarray} 
and
\begin{eqnarray}
&&U_C a^\dagger_h ({\bf p}) \vert 0>=
U_C a_h^\dagger U_C^{-1} \vert 0> = (-1)^{{1\over 2} +h}
b_{-h}^\dagger ({\bf p}) \vert 0>= \nonumber\\&&(-1)^{{1\over 2}+h} \vert
{\bf p}, -h >^-\,,\\
&&U_C b^\dagger_h ({\bf p}) \vert 0>= U_C
b_h^\dagger U_C^{-1} \vert 0> = (-1)^{{1\over
2}-h} a_{-h}^\dagger ({\bf p}) \vert 0>= \nonumber\\
&&(-1)^{{1\over
2}-h} \vert {\bf p}, -h >^+\,.
\end{eqnarray} 
Finally, for the $CP$ operation one should obtain: 
\begin{eqnarray}
&&U_P U_C a^\dagger_h ({\bf p}) \vert 0>=
-U_C U_P a^\dagger_h ({\bf p}) \vert 0> = (-1)^{{1\over 2}+h}
U_P b_{-h}^\dagger ({\bf p}) \vert 0> =\nonumber\\
&=& i (-1)^{{1\over 2} +h}
b_{h}^\dagger (-{\bf p}) \vert 0>= i (-1)^{{1\over 2}+h} \vert
-{\bf p}, h >^-\,,\\
&&U_P U_C b^\dagger_h ({\bf p}) \vert 0>= -U_C U_P b^\dagger_h
({\bf p}) = (-1)^{{1\over 2}-h} U_P
a_{-h}^\dagger ({\bf p})\vert 0> = \nonumber\\
&=&i (-1)^{{1\over
2}-h} a_{h}^\dagger (-{\bf p}) \vert 0>= i (-1)^{{1\over
2}-h} \vert -{\bf p}, h >^+\,.
\end{eqnarray} 
As in the classical case, the $P$ and $C$ operations anticommute
in the $({1\over 2},0)\oplus (0,{1\over 2})$ quantized case. This opposes
to the theory based on the 4-spinor eigenstates of chiral helicity
(cf.~\cite{DVO2,Nigam}).
Since the $V_T$ is an anti-unitary operator the problem must be solved
after taking into account of the fact that 
the $c$-numbers should be put outside
the hermitian conjugation
{\it without} complex conjugation:
\begin{equation}
[V_T h A V_T^{-1}]^\dagger = [h^\ast V_T A V_T^{-1} ]^\dagger
= h [V_T A^\dagger V_T^{-1} ]\,.
\end{equation}
After applying this definition we obtain:\footnote{$T$ should be  chosen 
in such a way in order to fulfill
$T^{-1} \gamma_0^T T= \gamma_0$, $T^{-1} \gamma_i^T T= \gamma_i$
and $T^T= -T$, as in Ref.~\cite{Bogol}.} 
\begin{eqnarray}
V_T a_h^\dagger ({\bf p}) V_T^{-1} &=& +i
(-1)^{{1\over 2}-h}
a_{h}^\dagger (-{\bf p})\,,\\
V_T b_h  ({\bf p}) V_T^{-1} &=& +i (-1)^{{1\over 2}-h}
b_{h} (-{\bf p})\,.
\end{eqnarray}
Furthermore, we observed that the parity properties depend on the phase factor
in the following definition:
\begin{equation}
U_P \Psi (t, {\bf x}) U_P^{-1} = e^{i\alpha} \gamma^0 \Psi (t, -{\bf
x})\,.  \label{def1}
\end{equation}
Indeed, 
\begin{eqnarray}
&&U_P a_h ({\bf p}) U_P^{-1} = -i e^{i\alpha} a_{-h} (-{\bf p})\,,\\
&&U_P b_h^\dagger ({\bf p}) U_P^{-1} = + i e^{i\alpha} b_{-h}^\dagger
(-{\bf p})\,.
\end{eqnarray} 
From this, if $\alpha=\pi/2$ we obtain
{\it opposite} parity properties of the creation/annihilation
operators for particles and anti-particles: 
\begin{eqnarray}
&&U_P a_h ({\bf p}) U_P^{-1} = + a_{-h} (-{\bf p})\,,\\
&&U_P b_h ({\bf p}) U_P^{-1} = - b_{-h}
(-{\bf p})\,.
\end{eqnarray}
However, the difference with the Dirac case still preserves
($h$ transforms to $-h$).  We find similar situation with 
the  question of constructing
the neutrino field operator (cf. with the Goldhaber-Kayser
creation phase factor).

Next, we find the explicit form of the parity operator
$U_P$ and prove that it commutes with the Hamiltonian operator.
We prefer to use the method described in~\cite[\S 10.2-10.3]{Greib}.
It is based on the {\it anzatz} that $U_P = \exp [i\alpha \hat A] \exp [i \hat
B]$ with $\hat A =\sum_{s}^{}\int d^3 {\bf p} [a_{{\bf p},s}^\dagger a_{-{\bf
p}s} +b_{{\bf p}s}^\dagger b_{-{\bf p}s}]$ and \linebreak $\hat B =\sum^{}_{s}\int d^3
{\bf p} [\beta a_{{\bf p},s}^\dagger a_{{\bf p}s} +\gamma b_{{\bf
p}s}^\dagger b_{{\bf p}s}]$. On using the known operator identity
\begin{equation}
e^{\hat A} \hat B e^{-\hat A} = \hat B +[\hat A,\hat B]_- +{1\over 2!}
[\hat A, [\hat A,\hat B]]+\ldots
\end{equation}
and $[\hat A,\hat B\hat C]_-= [\hat A,\hat B]_+ \hat C
-\hat B [\hat A,\hat C]_+$ one can fix  the parameters
$\alpha,\beta,\gamma$ such that one satisfies the physical
requirements that a Dirac particle and its anti-particle
have opposite intrinsic parities.

In our case, we  need to satisfy the requirement that  the operator
should invert not only the sign of the momentum, but the sign of
the helicity too. We may achieve this goal by the analogous postulate
$U_P= e^{i\alpha \hat A}$ with
\begin{equation}
\hat A =\sum_{h}^{} \int {d^3 {\bf p}\over 2E}
[a^\dagger_h ({\bf p}) a_{-h} (-{\bf p})
+b_h^\dagger ({\bf p}) b_{-h} (-{\bf p})]\,.
\end{equation}
By direct verification, the requirement
is satisfied provided that $\alpha=\pi/2$. You may compare this parity operator with
that given in~\cite{Itzyk,Greib} for Dirac fields:\footnote{Greiner used the following commutation
relations
$\left [ a ({\bf p}, s), a^\dagger ({\bf p}^\prime, s^\prime) \right ]_+ =
\left [ b ({\bf p}, s), b^\dagger ({\bf p}^\prime, s^\prime) \right ]_+ =
\delta^3 ({\bf p}-{\bf p}^\prime) \delta_{ss^\prime}$. One should also
note that the Greiner form of the parity operator is not the unique one.
Itzykson and Zuber~\cite{Itzyk} proposed another one differing by the
phase factors from (10.69) of~\cite{Greib}.
In order to find relations between those
two forms of the parity operator one should apply
additional rotation in the Fock space.}
\begin{eqnarray} \lefteqn{U_P =
\exp \left [ i{\pi\over 2} \int  d^3 {\bf p} \sum_s
\left ( a ({\bf p}, s)^\dagger
a (\tilde{\bf p},s) +b ({\bf p},s)^\dagger b (\tilde{\bf p},s)-
\right.\right.}\nonumber\\
&&\left.\left.- a ({\bf p},s)^\dagger a ({\bf p},s) + b ({\bf
p},s)^\dagger b ({\bf p},s) \right ) \right ]\,,\quad (10.69)\,
\mbox{of}\,
~\cite{Greib}.\end{eqnarray}
By
direct verification one can also come to the conclusion that  our new $U_P$ commutes
with the Hamiltonian:  \begin{equation} {\cal H} = \int d^3 {\bf x}
\Theta^{00} = \int d^3 {\bf k} \sum_h [ a_h^\dagger ({\bf k})
a_h ({\bf k}) - b_h ({\bf k}) b_h^\dagger ({\bf k})]\,,
\end{equation} i.e.
$[U_P, {\cal H} ]_- =0$\,.
Alternatively, we can try to choose other  commutation
relations~\cite{DVO2,DasG} for the set
of bi-orthonormal states. The formulas of the theory have been presented in the $({1\over
2},0)\oplus (0,{1\over 2})$ representation based on
the chiral helicity 4-eigenspinors, see below.  
Next, the theory, which is based on a system of 6-component Weinberg-like equations 
in the $(1,0)\oplus (0,1)$ representation, has also been constructed. 
The results are similar. 
The papers by Ziino and Barut~\cite{Barut}
and the Markov papers~\cite{Markov}  have connections
with the subject under consideration.


\section{Chiral Helicity Construct and the Different Definition of the Charge Conjugate Operator
on the Secondary Quantization Level.}

In the chiral
representation one can choose the spinorial basis (zero-momentum spinors)
in the following way:
\begin{eqnarray}
\lambda^S_\uparrow ({\bf 0})
&=& \sqrt{{m\over 2}}\begin{pmatrix}{0\cr i\cr 1\cr 0}\end{pmatrix},
\lambda^S_\downarrow ({\bf 0}) =
\sqrt{{m\over 2}}\begin{pmatrix}{-i\cr 0\cr 0\cr 1}\end{pmatrix},\\
\lambda^A_\uparrow ({\bf 0})
&=& \sqrt{{m\over 2}}\begin{pmatrix}{0\cr -i\cr 1\cr 0}\end{pmatrix},
\lambda^A_\downarrow ({\bf 0}) = \sqrt{{m\over 2}}
\begin{pmatrix}{i\cr 0\cr 0\cr 1}\end{pmatrix},
\end{eqnarray}
\begin{eqnarray}
\rho^S_\uparrow ({\bf 0})
&=& \sqrt{{m\over 2}}\begin{pmatrix}{1\cr 0\cr 0\cr -i}\end{pmatrix},
\rho^S_\downarrow ({\bf 0})
= \sqrt{{m\over 2}}\begin{pmatrix}{0\cr 1\cr i\cr 0}\end{pmatrix},\\
\rho^A_\uparrow ({\bf 0})
&=& \sqrt{{m\over 2}}\begin{pmatrix}{1\cr 0\cr 0\cr i}\end{pmatrix},
\rho^A_\downarrow ({\bf 0})
= \sqrt{{m\over 2}}\begin{pmatrix}{0\cr 1\cr -i\cr 0}\end{pmatrix}.
\end{eqnarray}

The indices $\uparrow\downarrow$ should be referred to the chiral helicity
quantum number introduced in Ref.~\cite{DasG}, $\eta =-h\gamma_5$. We use the notation of the
previous papers on the subject. Ahluwalia and Grumiller used 
the helicity basis for the 2nd-type 4-spinors.
The reader would immediately
find the 4-spinors of the second kind  $\lambda^{S,A}_{\uparrow\downarrow}
(p^\mu)$ and $\rho^{S,A}_{\uparrow\downarrow} (p^\mu)$
in an arbitrary frame on using the boost operators:
\begin{eqnarray}
\hspace{-5mm}\lambda^S_\uparrow (p^\mu) = \frac{1}{2\sqrt{E+m}}
\begin{pmatrix}{ip_l\cr i (p^- +m)\cr p^- +m\cr -p_r}\end{pmatrix},\,\,
\lambda^S_\downarrow (p^\mu)= \frac{1}{2\sqrt{E+m}}
\begin{pmatrix}{-i (p^+ +m)\cr -ip_r\cr -p_l\cr (p^+ +m)}\end{pmatrix},\\
\hspace{-5mm}\lambda^A_\uparrow (p^\mu) = \frac{1}{2\sqrt{E+m}}
\begin{pmatrix}{-ip_l\cr -i(p^- +m)\cr (p^- +m)\cr -p_r}\end{pmatrix},\,\,
\lambda^A_\downarrow (p^\mu) = \frac{1}{2\sqrt{E+m}}
\begin{pmatrix}{i(p^+ +m)\cr ip_r\cr -p_l\cr (p^+ +m)}\end{pmatrix},
\end{eqnarray}
and
\begin{eqnarray}
\hspace{-5mm}\rho^S_\uparrow (p^\mu) = \frac{1}{2\sqrt{E+m}}
\begin{pmatrix}{p^+ +m\cr p_r\cr ip_l\cr -i(p^+ +m)}\end{pmatrix},\,\,
\rho^S_\downarrow (p^\mu) = \frac{1}{2\sqrt{E+m}}
\begin{pmatrix}{p_l\cr (p^- +m)\cr i(p^- +m)\cr -ip_r} \end{pmatrix},\\
\hspace{-5mm}\rho^A_\uparrow (p^\mu) = \frac{1}{2\sqrt{E+m}}
\begin{pmatrix}{p^+ +m\cr p_r\cr -ip_l\cr i (p^+ +m)}\end{pmatrix},\,\,
\rho^A_\downarrow (p^\mu) = \frac{1}{2\sqrt{E+m}}
\begin{pmatrix}{p_l\cr (p^- +m)\cr -i(p^- +m)\cr ip_r}\end{pmatrix}\,.
\end{eqnarray}
Some of the 4-spinors are connected each other.
The normalization of the spinors $\lambda^{S,A}_{\uparrow\downarrow}
(p^\mu)$ and $\rho^{S,A}_{\uparrow\downarrow} (p^\mu)$ are the following ones:
\begin{eqnarray}
\overline \lambda^S_\uparrow (p^\mu) \lambda^S_\downarrow (p^\mu) \,&=&\,
- i m \quad,\quad
\overline \lambda^S_\downarrow (p^\mu) \lambda^S_\uparrow (p^\mu) \,= \,
+ i m \quad,\quad\\
\overline \lambda^A_\uparrow (p^\mu) \lambda^A_\downarrow (p^\mu) \,&=&\,
+ i m \quad,\quad
\overline \lambda^A_\downarrow (p^\mu) \lambda^A_\uparrow (p^\mu) \,=\,
- i m \quad,\quad\\
\overline \rho^S_\uparrow (p^\mu) \rho^S_\downarrow (p^\mu) \, &=&  \,
+ i m\quad,\quad
\overline \rho^S_\downarrow (p^\mu) \rho^S_\uparrow (p^\mu)  \, =  \,
- i m\quad,\quad\\
\overline \rho^A_\uparrow (p^\mu) \rho^A_\downarrow (p^\mu)  \,&=&\,
- i m\quad,\quad
\overline \rho^A_\downarrow (p^\mu) \rho^A_\uparrow (p^\mu) \,=\,
+ i m\quad.
\end{eqnarray}
All other conditions are equal to zero. 

Implying that $\lambda^S (p^\mu)$
(and $\rho^A (p^\mu)$) answer for positive-frequency solutions; $\lambda^A
(p^\mu)$ (and $\rho^S (p^\mu)$), for negative-frequency solutions, one can
deduce the dynamical coordinate-space equations~\cite{DVO2}:
\begin{eqnarray}
i \gamma^\mu \partial_\mu \lambda^S (x) - m \rho^A (x) &=& 0 \,,
\label{11}\\
i \gamma^\mu \partial_\mu \rho^A (x) - m \lambda^S (x) &=& 0 \,,
\label{12}\\
i \gamma^\mu \partial_\mu \lambda^A (x) + m \rho^S (x) &=& 0\,,
\label{13}\\
i \gamma^\mu \partial_\mu \rho^S (x) + m \lambda^A (x) &=& 0\,.
\label{14}
\end{eqnarray}
They can be written in the 8-component form. 
This is just  another representation of the Dirac-like equation in the appropriate Clifford Algebra.
One can also re-write the equations into the two-component form,
the Feynman-Gell-Mann equations.
In the Fock space the operators of the charge conjugation and space
inversions can be defined as in the previous Section. We imply the bi-orthonormal system of 
the anticommutation relations.
As a result we have the following properties of the creation (annihilation)
operators in the Fock space: 
\begin{eqnarray}
&&U^s_{[1/2]} a_\uparrow ({\bf p}) (U^s_{[1/2]})^{-1} = - ia_\downarrow
(-  {\bf p}),
U^s_{[1/2]} a_\downarrow ({\bf p}) (U^s_{[1/2]})^{-1} = + ia_\uparrow
(- {\bf p}),\\
&&U^s_{[1/2]} b_\uparrow^\dagger ({\bf p}) (U^s_{[1/2]})^{-1} =
+ i b_\downarrow^\dagger (- {\bf p}),
U^s_{[1/2]} b_\downarrow^\dagger ({\bf p}) (U^s_{[1/2]})^{-1} =
- i b_\uparrow (- {\bf p}),
\end{eqnarray}  
that signifies that the states created by the operators $a^\dagger
({\bf p})$ and $b^\dagger ({\bf p})$ have different properties
with respect to the space inversion operation, comparing with
Dirac states (the case also regarded in~\cite{Barut}):
\begin{eqnarray}
&&U^s_{[1/2]} \vert {\bf p},\,\uparrow >^+ = + i \vert -{\bf p},\,
\downarrow >^+,
U^s_{[1/2]} \vert {\bf p},\,\uparrow >^- = + i
\vert -{\bf p},\, \downarrow >^-,\\
&&U^s_{[1/2]} \vert {\bf p},\,\downarrow >^+ = - i \vert -{\bf p},\,
\uparrow >^+,
U^s_{[1/2]} \vert {\bf p},\,\downarrow >^- =  - i
\vert -{\bf p},\, \uparrow >^-.
\end{eqnarray}

For the charge conjugation operation in the Fock space we have
two physically different possibilities. The first one, {\it e.g.},
\begin{eqnarray}
&&U^c_{[1/2]} a_\uparrow ({\bf p}) (U^c_{[1/2]})^{-1} = + b_\uparrow
({\bf p}),
U^c_{[1/2]} a_\downarrow ({\bf p}) (U^c_{[1/2]})^{-1} = + b_\downarrow
({\bf p}),\nonumber\\
&&\\
&&U^c_{[1/2]} b_\uparrow^\dagger ({\bf p}) (U^c_{[1/2]})^{-1} =
-a_\uparrow^\dagger ({\bf p}),
U^c_{[1/2]} b_\downarrow^\dagger ({\bf p})
(U^c_{[1/2]})^{-1} = -a_\downarrow^\dagger ({\bf p}),\nonumber\\
\end{eqnarray}
in fact, has some similarities with the Dirac construct.
The action of this operator on the physical states are
\begin{eqnarray}
U^c_{[1/2]} \vert {\bf p}, \, \uparrow >^+ &=& + \,\vert {\bf p},\,
\uparrow >^- \,,\,
U^c_{[1/2]} \vert {\bf p}, \, \downarrow >^+ = + \, \vert {\bf p},\,
\downarrow >^- ,\\
U^c_{[1/2]} \vert {\bf p}, \, \uparrow >^-
&=&  - \, \vert {\bf p},\, \uparrow >^+ ,\,
U^c_{[1/2]} \vert
{\bf p}, \, \downarrow >^- = - \, \vert {\bf p},\, \downarrow >^+ .
\end{eqnarray}  
But, one can also construct the charge conjugation operator in the
Fock space which acts, {\it e.g.}, in the following way:
\begin{eqnarray}
&&\widetilde U^c_{[1/2]} a_\uparrow ({\bf p}) (\widetilde U^c_{[1/2]})^{-1}
= - b_\downarrow ({\bf p}), \widetilde U^c_{[1/2]}
a_\downarrow ({\bf p}) (\widetilde U^c_{[1/2]})^{-1} = - b_\uparrow
({\bf p}),\\
&&\widetilde U^c_{[1/2]} b_\uparrow^\dagger ({\bf p})
(\widetilde U^c_{[1/2]})^{-1} = + a_\downarrow^\dagger ({\bf
p}),
\widetilde U^c_{[1/2]} b_\downarrow^\dagger ({\bf p})
(\widetilde U^c_{[1/2]})^{-1} = + a_\uparrow^\dagger ({\bf p}).
\end{eqnarray}
Therefore, 
\begin{eqnarray}
\widetilde U^c_{[1/2]} \vert {\bf p}, \, \uparrow >^+ &=& - \,\vert {\bf
p},\, \downarrow >^- ,\,
\widetilde U^c_{[1/2]} \vert {\bf p}, \, \downarrow
>^+ = - \, \vert {\bf p},\, \uparrow >^- ,\\
\widetilde U^c_{[1/2]} \vert
{\bf p}, \, \uparrow >^- &=& + \, \vert {\bf p},\, \downarrow >^+
,\,
\widetilde U^c_{[1/2]} \vert {\bf p}, \, \downarrow >^- = + \, \vert {\bf
p},\, \uparrow >^+ .
\end{eqnarray}

Next, by straightforward
verification one can convince ourselves about correctness of the
assertions made in~[11b] (see also~\cite{Nigam}) that it is
possible to have a situation when the operators of the space inversion and
the charge conjugation commute each other in the Fock space. For instance,
\begin{eqnarray}
U^c_{[1/2]} U^s_{[1/2]} \vert {\bf
p},\, \uparrow >^+ &=& + i U^c_{[1/2]}\vert -{\bf p},\, \downarrow >^+ =
+ i \vert -{\bf p},\, \downarrow >^- \\
U^s_{[1/2]} U^c_{[1/2]} \vert {\bf
p},\, \uparrow >^+ &=& U^s_{[1/2]}\vert {\bf p},\, \uparrow >^- = + i
\vert -{\bf p},\, \downarrow >^- .
\end{eqnarray}  
The second choice of the charge conjugation operator answers for the case
when the $\widetilde U^c_{[1/2]}$ and $U^s_{[1/2]}$ operations
anticommute:
\begin{eqnarray}
\widetilde U^c_{[1/2]} U^s_{[1/2]} \vert {\bf p},\, \uparrow >^+ &=&
+ i \widetilde U^c_{[1/2]}\vert -{\bf
p},\, \downarrow >^+ = -i \, \vert -{\bf p},\, \uparrow >^- \,\,\\
U^s_{[1/2]} \widetilde U^c_{[1/2]} \vert {\bf p},\, \uparrow >^+ &=& -
U^s_{[1/2]}\vert {\bf p},\, \downarrow >^- = + i \, \vert -{\bf p},\,
\uparrow >^- .
\end{eqnarray}  

Finally, the time reversal {\it anti-unitary} operator in
the Fock space should be defined in such a way that the formalism to be
compatible with the $CPT$ theorem. If we wish the Dirac states to transform
as $V(T) \vert {\bf p}, \pm 1/2 > = \pm \,\vert -{\bf p}, \mp 1/2 >$ we
have to choose (within a phase factor), Refs.~\cite{Itzyk,Bogol}:
\begin{equation}
S(T) = \begin{pmatrix}{\Theta_{[1/2]} &0\cr 0 &
\Theta_{[1/2]}\cr} \end{pmatrix}\quad.
\end{equation}
Thus, in the first relevant case we obtain for the $\Psi
(x^\mu)$ field, composed of $\lambda^{S,A}$ or $\rho^{A,S}$ 4-spinors
\begin{eqnarray}
V^{^T} a^\dagger_\uparrow ({\bf p}) (V^{^T})^{-1} &=& a^\dagger_\downarrow
(-{\bf p}),\,\,
V^{^T} a^\dagger_\downarrow ({\bf p}) (V^{^T})^{-1} = -
a^\dagger_\uparrow (-{\bf p}) \nonumber\\
&&\\
V^{^T} b_\uparrow ({\bf p}) (V^{^T})^{-1} &=& b_\downarrow
(-{\bf p}),\,\,
V^{^T} b_\downarrow ({\bf p}) (V^{^T})^{-1} = -
b_\uparrow (-{\bf p}),\nonumber\\
\end{eqnarray} 
that is not surprising. 

\section{The Conclusions.}

Thus,  we proceeded as in the textbooks and defined the parity matrix as $P=\gamma_0$, 
because the helicity 4-spinors satisfy the Dirac equation, and the 2nd-type $\lambda$-spinors satisfy the coupled Dirac equations. Nevertheless, the properties of the corresponding spinors appear to be different with respect to the parity both
on the first and the second quantization level. This result is compatible with the statement made by Berestetskii, Lifshitz and Pitaevskii. We  defined another charge conjugation operator in the Fock space, which also transforms the positive-energy 4-spinors to the negative-energy ones. In this case the operators
$P$ and $C$ commute (instead of the anticommutation in the Dirac case), that is related to the eigenvalues of the charge operator, as in the Foldy and Nigam paper.


\end{document}